# Fast ion diffusion, superionic conductivity and phase transitions of the nuclear materials $UO_2$ and $Li_2O$.

Prabhatasree Goel, N. Choudhury and S. L. Chaplot
Solid State Physics Division
Bhabha Atomic Research Centre, Mumbai 400 085, India

**Abstract**

Lattice dynamics and molecular dynamics studies of the oxides $UO_2$ and $Li_2O$ in their normal as well as superionic phase are reported. Lattice dynamics calculations have been carried out using a shell model in the quasiharmonic approximation. The calculated elastic constants, phonon frequencies and specific heat are in good agreement with reported experimental data, which help validate the interatomic potentials required for undertaking molecular dynamics simulations. The calculated free energies reveal high pressure fluorite to cottunite phase transitions at 70 GPa for $UO_2$ and anti-fluorite to anti-cotunnite phase transformation at 25 GPa for $Li_2O$, in agreement with reported experiments. Molecular dynamics studies shed important insights into the mechanisms of diffusion and superionic behavior at high temperatures. The calculated superionic transition temperature of $Li_2O$ is 1000 K, while that of $UO_2$ is 2300 K.

## 1. Introduction

$UO_2$ is of technological importance owing to its use as a nuclear fuel [1]. Knowledge of the thermodynamic and transport properties of nuclear materials [1-18] at high temperatures is of great interest. $UO_2$ belongs to the class of superionics, wherein fast-ion conduction processes involving rapid diffusion of a significant fraction of the oxygen atoms within an essentially rigid framework of uranium atoms occurs. Microscopic modeling or simulation is necessary to understand the conduction processes and thermodynamic properties at high temperatures and pressures of superionic crystals. $UO_2$ has a face-centered-cubic fluorite structure having space group $O_h^5$ (*Fm3m*), with the oxygen atoms in the tetrahedral sites. $UO_2$ and $Li_2O$ show a type II superionic transition [1,10] attaining high levels of ionic conductivity following a gradual and continuous disordering process within the same phase.

Several theoretical and experimental works [4, 19-30] have been reported on numerous fast-ion conductors like $Li_2O$, $CaF_2$, $BaF_2$, $PbF_2$, $SrCl_2$, $CuI$, *etc*. The main impetus for these studies has been to unravel the causes behind the process of fast-ion conduction. In the case of $Li_2O$, further interest has been to study Li diffusion from the point of view of tritium generation for future fusion reactors. The oxides $Li_2O$ and $UO_2$ behave similar to other superionic halides. The extensive diffusion is characterized by a large decrease in the elastic constant $C_{11}$ and specific heat anomaly at the transition temperature $T_c$ [28-33]. Neutron scattering measurements [34] indicate that the anionic sub-lattice in $UO_2$ becomes heavily disordered in the region of 2300 K. Measured elastic constants [35] do show a softening above 2400 K in the region where fast-ion behavior is expected in $UO_2$, but the variation below this temperature is already very large. There is a large increase in specific heat [36-38]] at high temperatures in $UO_2$. $Li_2O$ shows a sudden decrease in the value of the $C_{11}$ elastic constant at the transition temperature, $T_c \sim 1200$ K (the melting point $T_m$ of $Li_2O$ is 1705 K [22]), but there seems no drastic change in the specific heat [39,40]. Both these compounds comply with the general belief that fluorites (anti-fluorites) in general show a diffuse transition at about $0.8T_m$ ($T_m$ = melting point, $T_m$ of $UO_2$ is 3120 K). Above the transition temperature the diffusion coefficient of one of the constituent atom becomes comparable to that of liquids. Detailed study of the processes occurring in the crystal lattice at elevated temperatures is essential to understand the transitions.

Angle dispersive synchrotron X-ray powder diffraction and Raman spectroscopy experiments reveal a reversible phase transition from cubic anti-fluorite to the orthorhombic anti-cotunnite structure at a pressure near 50 GPa for $Li_2O$ [41-43]. This transition is accompanied by a relatively large volume collapse of about 5.4($\pm$0.8)% and a large hysteresis upon pressure reversal ($P_{down}$ at ~25 GPa). Similarly, $UO_2$ also shows a sluggish transformation to cotunnite-type phase at about 40 GPa, the cotunnite phase coexists with the fluorite phase even at 69 GPa. [44,45].

The present study is aimed at formulating a suitable interatomic potential to explain the vibrational properties of the oxides in concurrence with the available experimental data, as in our previous work [4]. The main objectives of the present study are: (i) to determine a suitable interatomic potential model to calculate the phonon spectrum, specific heat, other thermodynamic and elastic properties, (ii) to carry out molecular dynamics simulations using these interatomic potentials to elucidate diffusion behavior and the thermodynamic properties of the oxides at elevated temperatures, and (iii) to study the phase transformation of the fluorite (anti-fluorite) to cotunnite (anti-cotunnite) phase.

## 2. Lattice dynamics calculations and molecular dynamics calculations

Our calculations have been carried out in the quasiharmonic [46-49] approximation using the interatomic potentials consisting of Coulomb and short-range Born-Mayer type interaction terms:

$$V(r_{ij}) = \frac{e^2}{4\pi\varepsilon_0} \frac{Z(k)Z(k')}{r_{ij}^2} + a \exp\left[\frac{-br_{ij}}{R(k)+R(k')}\right] \qquad (1)$$

where, $r_{ij}$ is the separation between the atoms ? and ? of type $k$ and $k'$ respectively. $R(k)$ and $Z(k)$ are the effective radius and charge of the $k^{th}$ atom, $a$ and $b$ are the empirical parameters optimized from several previous calculations [48,49]. Oxygen atoms have been modeled using a shell model [46,47], where a mass less shell is linked to the atomic core of charge $Y(k)$ by harmonic force constant $K(k)$. Lattice constant, zone center phonon frequencies and elastic constants have been fitted to experimental values. The calculations have been carried out using the current version of the software DISPR developed in Trombay [50,51]. The interatomic potential enables the calculation of the phonon frequencies in the entire Brillouin zone. Based on the crystal symmetry, group theoretical analysis provides a classification of the frequencies at zone center and the symmetry directions, in the various representations.

Molecular dynamics is a powerful method for exploring the structure and dynamics of solids, liquids and gases. Explicit computer simulation of the structure and dynamics using this technique allows a microscopic insight into the behavior of materials to understand the macroscopic phenomenon like diffusion of lithium (oxygen in case of $UO_2$) ions and their contribution to the fast ion transit ion in this case. An interatomic potential, which treats Li, U and O as rigid units may be sufficient to study properties like diffusion. The optimized parameters obtained from lattice dynamics studies have been used for these simulations. In our study, we have taken a macro cell of a large number of rigid atoms with periodic boundary conditions to study the response of the system when set free to evolve from a configuration disturbed from the equilibrium situation. The lattice parameters and atomic trajectories can thus be obtained as a function of temperature and external pressure. Calculations in this work have been done using the software developed at Trombay [51-54]. The simulations have been done at various temperatures up to and beyond the fast ion transition. In our study we have considered a macro cell of 768 rigid atoms with periodic boundary conditions in case of $Li_2O$ and 1500 rigid atoms in case of $UO_2$.

## 3. Results and Discussion

### 3.1. Phonon spectra and elastic properties

The calculated values of the lattice parameter, bulk modulus, and elastic constants compare well with the experimentally obtained data as given in Table I. The computed phonon dispersion relations in $Li_2O$ [4] and $UO_2$ along the various high symmetry directions are in good agreement with available experimental data [55, 56]. The elastic behavior of the two oxides is markedly different (Table I). $UO_2$ is a harder material with almost twice the value of bulk modulus as compared to $Li_2O$.

The total and partial densities of $Li_2O$ [4] and $UO_2$ are given in Fig. 2. In case of $Li_2O$, the energy spans the spectral range up to 90 meV, while for $UO_2$ it is up to 75 meV. From the partial densities of states, we conclude that Li atoms in $Li_2O$ contribute almost in the entire range upto 75 meV with significant contribution at 90 meV as well. Uranium's contribution is restricted up to 25 meV only. The diffusing atom Li in $Li_2O$ shows a behavior similar to the one exhibited by oxygen in $UO_2$, but owing to its large mass, uranium's behavior is clearly opposite to that of the non-diffusing oxygen in $Li_2O$. The oxygens contribute over the entire energy range, although their spectra are different in $Li_2O$ and $UO_2$.

### 3.2. Specific heat

The calculated density of states has been used to evaluate various thermodynamic properties of the two oxides. Calculated specific heat at constant pressure, $C_P(T)$ have been compared with available data [36-40] in Fig. 3 for both the systems. In $Li_2O$, the comparison is very good up to 1100 K beyond which the fast-ion behavior sets in and the slope of the experimental data [39] is much greater compared to the calculations [4]. We have incorporated the anharmonic corrections from the implicit effects involving volume thermal expansion in the quasiharmonic lattice dynamics calculations. For temperatures above T=1100 K, in $Li_2O$ explicit anharmonic effects involving contributions from higher order terms of the crystal potential become important, this gives rise to the disagreement between the lattice dynamics calculations and the reported data.

Using 96-atom supercells, we have estimated the Frenkel defect energies ($E_F$) defined as the energy required for the formation of a vacancy/interstitial atom pair. Our calculated $E_F$ values (Table II) are in satisfactory agreement with reported first principles calculations [14,33] and experimental data [15,19]. As report ed by various workers [16,17, 36], defects are not believed to contribute significantly to the observed specific heat $C_P(T)$ in $UO_2$ and $Li_2O$ in the 0-1600 K temperature range reported in the present study. In case of $UO_2$ [36], in addition to the disordering of the oxygen sub-lattice, there are various other factors like electronic excitations, valence-conduction band transitions, *etc.* which play a significant role in the anomalous increase in the specific heat which sets in above T=1600 K [16], well before the fast-ion transition. Hence the disagreement between computed and experimental specific heats is higher in case of $UO_2$ as can be seen in fig. 3.

### 3.3. Molecular dynamics results

The diffusion coefficient of the two oxides have been calculated from 300K to 1500 K incase of $Li_2O$, and up to 3000 K in case of $UO_2$ using molecular dynamics simulations (Fig. 4). The diffusion coefficient of Li [4] has been compared with available experimental data. The diffusion coefficient is comparable to that of a liquid in the superionic phase. Both the oxides show fast-ion conduction as expected. Our molecular dynamics results suggest that the superionic phase sets in around 1000 K in case of $Li_2O$ while in $UO_2$ it sets in around 2300 K. Superionic conductivity is a complex phenomena and the computed transition temperature (T=1000 K) in $Li_2O$ can be regarded as being in good agreement with the observed fast-ion transition temperature of around 1200 K [57]. The signature of a corresponding superionic transition is found indirectly in the enthalpy studies on $UO_2$, since direct measurements are made difficult with high temperatures involved. It is found to undergo a bredig transition (involving jump in specific heat across the normal to superionic phase transition) at about 2610 K [16,17]. To the best of our knowledge, there are no available experimental studies on the diffusion coefficient of oxygen ions in $UO_2$.

### 3.4. Phase transformations

These oxides are found to undergo pressure induced transformations to orthorhombic structures. Anti-fluorite lithium oxide undergoes a transition to the anti-cotunnite phase at pressures of about 50 GPa [41,42], this transition is accompanied by a relatively large volume collapse of ~5.4(0,8)% and a large hysteresis upon pressure reversal, while decreasing transition value of the pressure is found to be about 25 GPa.

Fig. 5 gives the calculated free energy of the two phases with increasing pressure which reveal a free energy crossover and an anti-fluorite to anti-cottunite phase transition at about 25 GPa in $Li_2O$. The ratio of the volume of the anti-cotunnite phase at this pressure with respect to the corresponding volume of the anti-fluorite at the same pressure is about 6%. In the case of $UO_2$, reported experimental studies [44,45] reveal a sluggish transformation, wherein the cotunnite phase first appears at about 40 GPa and the fluorite phase is found to coexist even at 69 GPa. Our calculations (Fig. 5) show the transition point to be 70 GPa, with a volume decrease of about 3.5 % with respect to the fluorite volume. This behavior is in accordance to the structural variation of other superionic compounds with pressure.

### 4. Conclusions

Lattice dynamics calculations of the vibrational and thermodynamic properties of $Li_2O$ and $UO_2$ have been carried out using shell models. The elastic constants, bulk modulus, equilibrium lattice constant and phonon frequencies are in very good agreement with reported data. Both the oxides show a transition to the fast-ion phase at elevated temperatures. MD simulations reveal that $Li_2O$ becomes superionic at around 1000 K while $UO_2$ shows a transition at around 2300 K. Diffusion coefficients at temperatures $T\sim 0.8T_m$, are comparable to that of liquids. As reported in the literature [41,42], $Li_2O$ shows a transition to anti-cotunnite phase at around 25 GPa. $UO_2$ undergoes similar transformation at a higher pressure of 70 GPa.

**Table I** Comparison between the calculated and experimental lattice parameters, elastic constants of $Li_2O$ and $UO_2$.

| Physical Quantity | Calc. $Li_2O$ | Expt $Li_2O$[2,39] | Calc. $UO_2$ | Expt. $UO_2$[34] |
|---|---|---|---|---|
| Lattice Parameter (nm) | 0.461 | 0.46 | 0.546 | 0.547 |
| Bulk Modulus (GPa) | 103 | 82 | 180.5 | 207 |
| $C_{11}$ (GPa) | 213 | 202 | 387 | 389 |
| $C_{44}$ (GPa) | 52 | 59 | 66 | 60 |
| $C_{12}$ (GPa) | 56 | 21 | 77 | 119 |

Table II. Comparison of the calculated Frenkel defect formation energies (eV) with reported first principles [14,33] and atomistic calculations [18] and experimental data [14-15,19]. While in $Li_2O$, it involves cation (Li) vacancy/interstitial pair formation, it involves the anions (O) in $UO_2$.

|  | This work $E_F$ | *Ab initio* calculations $E_F$ | Atomistic simulations $E_F$ | Experimental $E_F$ |
|---|---|---|---|---|
| $Li_2O$ | 2.0 | 2.2 [33] |  | 1.58-2.53 [19] |
| $UO_2$ | 4.1 | 3.9 [14] | 5.4 [18] | 4.6 ± 0.5 [58,59], 3.0-4.6 [14,15] |

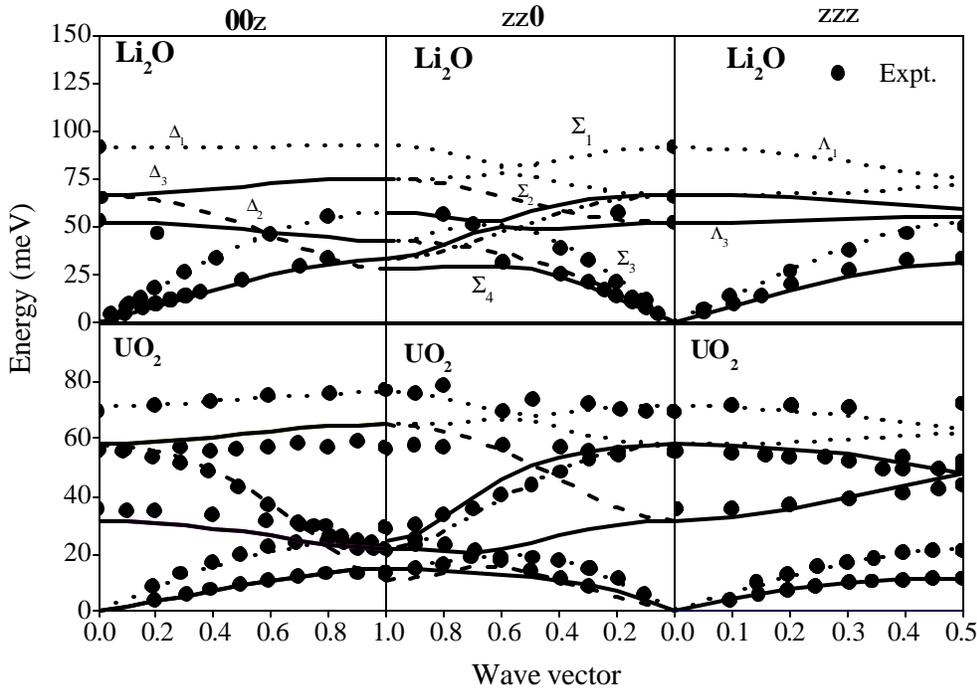

Figure 1: Comparison of the calculated (full, dashed and dash-dot lines) phonon dispersion relations with experimental (symbols) neutron scattering data in $Li_2O$ [55] and $UO_2$ [56].

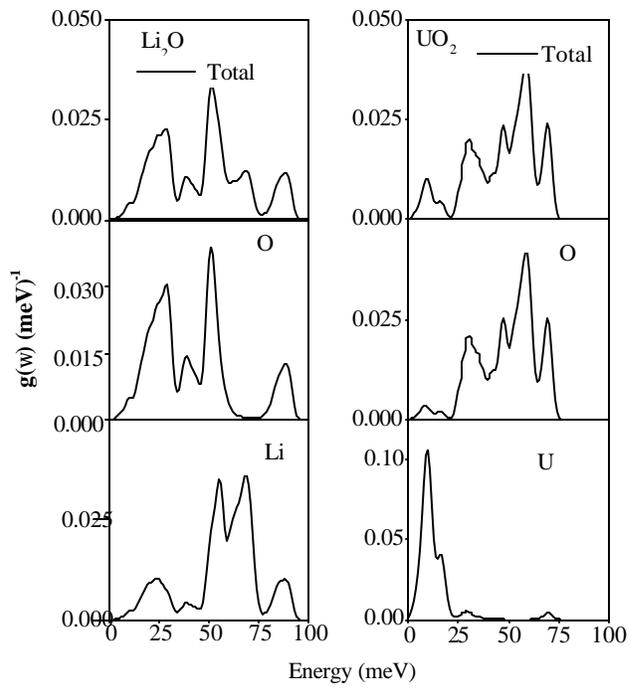

Figure 2. Phonon density of states along with the partial density of states for lithium, oxygen and uranium in Li$_2$O and UO$_2$ as calculated by quasiharmonic lattice dynamics calculations.

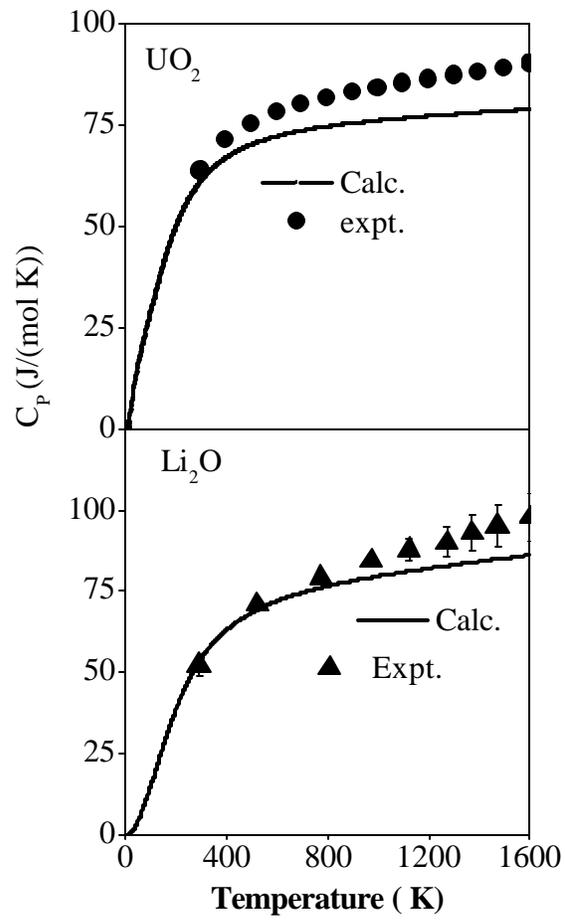

Figure 3. Specific heat at constant pressure compared with experimental data (closed symbols) for $UO_2$ [36] and $Li_2O$ [39].

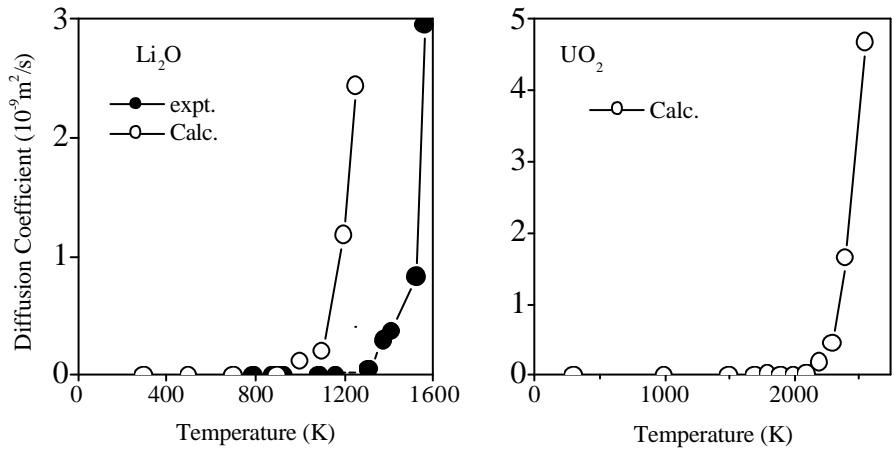

Figure 4. Diffusion coefficient of Li in $Li_2O$ and O in $UO_2$, as a function of temperature. Open circles are the calculated values while closed circles are the experimental [57] values as taken from the literature.

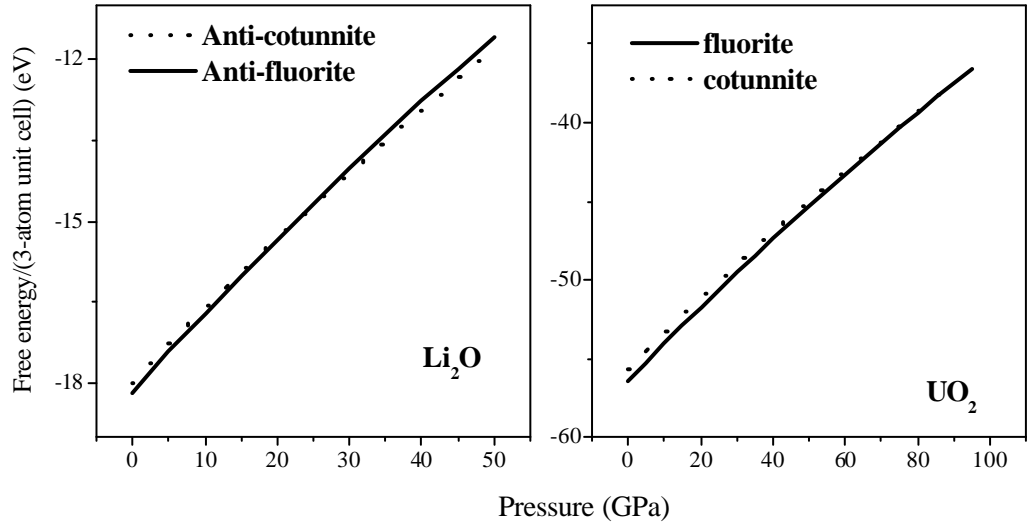

Figure 5. Calculated free energies of $Li_2O$ ($UO_2$) which reveal anti-fluorite to anti-cottunite (fluorite to cottunite) phase transitions at pressures of 25 (70) GPa